\documentclass[journal, printdates]{rmaa}

\usepackage{graphicx}	
\usepackage{paralist}
\usepackage[utf8]{inputenc}
\usepackage{color} 
\usepackage{soul}
\usepackage{float}
\usepackage[caption=false]{subfig}

\newcommand{\hMsun}{$h^{-1}\rm{M_{\odot}}$}

\title{Substructures in Minor Mergers' Tidal Streams}

\author{
  D.A. Nore\~na,\altaffilmark{1\dag} 
  J. C. Mu\~noz-Cuartas,\altaffilmark{1}
  L.F. Quiroga, ,\altaffilmark{1}
  and N. Libeskind\altaffilmark{2,3}}

\altaffiltext{1}{FACom - Instituto de F\'isica, FCEN, Universidad de
  Antioquia (UdeA), Calle 70 No. 52-21, Medell\'in, Colombia.}

\altaffiltext{2}{Leibniz-Institut f\"ur Astrophysik Potsdam (AIP), An
  der Sternwarte 16, D-14482 Potsdam, Germany.}

\altaffiltext{3}{l'Institut de Physique Nucl\'eaire de Lyon (IPNL),
  University of Lyon; UCB Lyon 1/CNRS/IN2P3; Lyon, France.}

\shortauthor{Nore\~na, Mu\~noz-Cuartas, Quiroga, Libeskind}
\shorttitle{Substructures in Tidal Streams}

\fulladdresses{

\item FACom - Instituto de F\'isica, FCEN, Universidad de Antioquia
  (UdeA), Calle 70 No. 52-21, Medell\'in, Colombia. (\dag
  david.norena@udea.edu.co).
  
\item Leibniz-Institut f\"ur Astrophysik Potsdam (AIP), An der
  Sternwarte 16, D-14482 Potsdam, Germany.
  
\item l'Institut de Physique Nucl\'eaire de Lyon (IPNL), University of
  Lyon; UCB Lyon 1/CNRS/IN2P3; Lyon, France.}

\listofauthors{D.A. Nore\~na\dag , J. C. Mu\~noz-Cuartas,  L.F. Quiroga, N. Libeskind}
\indexauthor{D.A. Nore\~na}
\indexauthor{J.C. Mu\~noz-Cuartas}
\indexauthor{L.F. Quiroga}
\indexauthor{N. Libeskind}

\resumen{En este trabajo se explora si subestructuras como cúmulos
  estelares se pueden formar del puente de marea producido durante un
  minor merger. Usamos simulaciones de N-cuerpos y SPH de una galaxia
  satélite interactuando con su galaxia host. Estudiamos la
  distribución de masa en los streams para identificar sobredensidades
  en las que se podrían formar subestructuras. Encontramos que sin gas
  no se da formación de subestructuras pues ninguna sobredensidad
  muestra morfología definida ni estabilidad dinámica. Incluyendo gas
  encontramos la formación de muchos grumos, estructuras longevas
  físicamente ligadas ($t$ $\geq$ 1 Gyr). Analizamos las órbitas,
  edades y masas de estas estructuras, encontrando una correspondencia
  con los subsistemas del halo. Concluimos que es posible formar
  estructuras como cúmulos estelares del material disponible en los
  puentes de marea y encontramos evidencia en favor de la presencia de
  materia oscura en esos sistemas.}

\abstract{In this work, we explore the idea that substructures like
  stellar clusters could be formed from the tidal stream produced in
  galactic minor mergers. We use $N$-body and SPH simulations of
  satellite galaxies interacting with a larger galaxy. We study the
  distribution of mass in streams to identify overdensity regions in
  which a substructure could be formed. We found that without gas, no
  substructure formed as none of the overdensities shows a definite
  morphology nor dynamical stability. Including gas we found that
  several clumps appear and proved to be real long standing physical
  structures ($t \geq$ 1 Gyr). We analyzed the orbits, ages and masses
  of these structures, finding its correspondence with the halo
  subsystems. We conclude that it is possible to form cluster-like
  structures from the material in tidal streams and found evidence in
  favour of the presence of dark matter in these systems.}

\addkeyword{galaxies: interactions}
\addkeyword{galaxies: evolution}
\addkeyword{galaxies: star clusters}
\addkeyword{(Galaxy:) globular clusters: general}
\addkeyword{(Galaxy:) open clusters and associations: general}
\addkeyword{methods: numerical}

\begin{document}
\maketitle


\section{Introduction}
\label{sec:intro}
The galactic halo has plenty of astrophysical systems evolving under
the interaction of the different galactic components. These
substructures have diverse nature, dynamics and origins and together
constitute the building blocks of the ongoing galaxy formation
process. Among others, there are many stellar subsystems as the open
and globular clusters~\citep{Binney2008}, pure gaseous ones as high
velocity clouds (HVC)~\citep{Wakker1997} and combined gaseous and
stellar systems such as tidal streams and satellite
galaxies~\citep{Ibata2001}.

Open and globular clusters are segregated by several
characteristics. Open clusters are considered as more young, metal
rich than their globular counterparts, in addition, open clusters are
associated spatially with the galactic disc while globular are mostly
distributed spherically all around the halo. This segregation suggest
that their formation processes are diverse. In one hand, the formation
of open clusters is considered well understood as the collapse and
fragmentation of molecular clouds in the galactic
disk~\citep{Elmegreen1997}. The case of globular clusters exhibits a
greater degree of complexity because actually there are two
subpopulations of them. There is a metal poor globular cluster
population (MPGC) extended across the halo, and the young, metal rich
population (MRGC)~\citep{Carroll2006}. In addition, there are several
cases that do not fit very well in the two previous subpopulations as
it is the case of the globular cluster $\omega$-Centauri (henceforth
$\omega$-Cen) mainly due to its unusual size and metallicity
dispersion~\citep{Harris1999}. This variety suggests that even only
for the globular clusters there are diverse formation mechanisms.\\
Different models have been proposed to explain possible formation
mechanisms for the two subpopulations of globular clusters in The
Galaxy. For the old MPGC subpopulation the widely accepted hypothesis
is that they come from primordial density fluctuations in the density
field at very high redshift, when the universe expanded and cooled to
a temperature of about 4000K and the baryonic density was
approximately $10^4$ atoms cm${}^{-3}$ \citep{Reina-Campos2019}. Under
this conditions, the only density fluctuations that can grow with time
has wavelength in excess of the critical Jeans length of about 5
pc~\citep{Peebles1968}. \\
For the young MRGC subpopulation, several models have been proposed
but it appears that there is not a single mechanism that can form all
existing MRGC in a given galaxy \citep{Ashman1992, Bekki2003,
  Shapiro2010}. One of the main models suggests that a significant
fraction of the metal-rich subpopulation may have originated in
interacting galaxies, both minor and major
mergers~\citep{Ashman1992}. Major mergers cause several starburst
episodes in the gaseous component of each galaxy, and globular
clusters can be formed in regions with high gas
density~\citep{Li2004}. Minor mergers may also contribute to the young
population with clusters formed within the small satellite galaxy from
the interaction with the larger galaxy~\citep{Zepf1993}. Also, the
globular cluster system of the minor galaxy would eventually be
accreted by the largest galaxy, also contributing to the MPGCs
subpopulation~\citep{Forbes2010}. The minor merger scenario can be
seen in the Magellanic Clouds, where there is observational evidence
of ongoing cluster formation and an ancient cluster system bound to
the clouds~\citep{Harris1998, Georgiev2010}. It was further suggested
that the very central region of a satellite galaxy could form a
globular cluster as the bound structure surviving the effects of the
tidal stripping induced by its host galaxy ~\citep{Bekki2002}.\\
Moreover, recently observational evidence that suggests that several
(if not all) GCs contain various stellar populations has come to
light. For example, many GC stars have the same amount of Fe (and
other heavy elements) inside a specific radius, but a wide variation
in light elemental abundance (Li-Ai) on a star-to-star basis
\citep{Conroy2011}. \citet{Norris2011} is a crucial study in this
problem; they showed that some ultra-compact dwarf galaxies have color
magnitude diagrams indistinguishable from those of GCs and the nuclei
of dwarf galaxies. \citet{Bekki2003} found that the multiple stellar
populations of $\omega$-Cen can be explained in terms of a nucleated
dwarf galaxy scenario: the tidal field of the host galaxy induces gas
inflow towards the center of the cluster progenitor, triggering
multiple star bursts that lead to chemical enrichment. Other GC
candidates that are thought to have formed in ostensibly dark matter
potential wells deep enough to retain self-enriched Fe produced by
supernovae Ia explosions include M22, NGC 1851 and Andromeda's G1. The
evidence showing chemical complexity of the cluster stellar
populations suggest that the classical picture of all GC's belonging
to a single monolithic population should be reevaluated. \\
Similarly, HVCs appear to be the result of two possible mechanisms:
One is the return to the disc of gas and dust expelled via supernovae
events and the other is the infall of gas and dust from a stripped
subsystem, such as globular clusters or satellite
galaxies~\citep{Wakker1997}. \\
A combination of both processes is necessary to explain the current
distribution of high and intermediate velocity clouds. For example,
from hydrodynamical simulations it is concluded that most massive HVC
such as the well known Complex C were originated from ejection of
material from the Milky Way's disc \citep{Fraternali2015}; but the
velocity dispersion, the metallicity, sizes and masses of the smallest
clouds are consistent with an extragalactic origin \citep{Blitz1999,
  Binney2009}.\\
\begin{figure*}[h]
  \begin{center}
    \hspace*{-1cm}\includegraphics[width=0.9\textwidth]{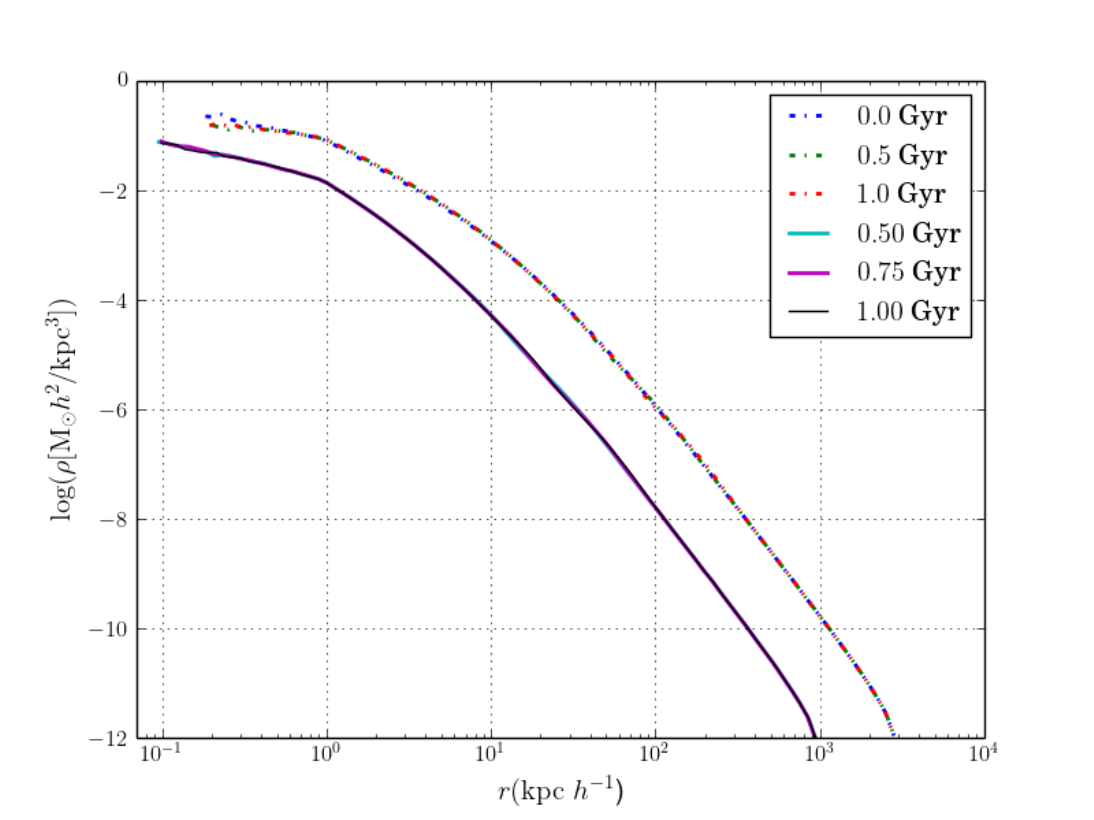}
    \caption{Convergence of the density profiles of the host (top) and
      satellite (bottom) dark matter halos under numerical
      relaxation. The galaxies were evolved isolated for 1 Gyr.}
    \label{fig:NumRelax}
  \end{center}
\end{figure*}
As a result of the tidal disruption of the Galaxy subsystems, the
so-called tidal streams are originated. They are composed in most
cases by stars and gas \citep{Belokurov2006}, like the Magellanic
Stream, where recent observations have confirmed the presence of a
young open star cluster most likely formed in the stream
~\citep{Price-Whelan2018}. Remarkably, all the streams observed in the
Milky Way galaxy are clearly nohomogeneous and exhibit overdensities
~\citep{Kupper2012}. These overdensities evolve in the galactic
potential as well, undergoing different processes that could
eventually transform them into self gravitating systems like
clusters. The main purpose of this work is to determine through
$N-$body simulations of galaxy minor mergers if the overdensities in
the tidal streams could really meet the conditions to be considered
self-gravitating substructures. In a future work, we will investigate
under what conditions the evolution of such substructures could lead
to the formation of real astrophysical systems such as globular
clusters and high velocity clouds.\\
This paper is organised as follows: In Section 2 we describe the whole
setup of the $N-$body simulations, from the determination of the
satellite galaxy initial position to the structure of the host galaxy
passing through the astrophysical characteristics of the satellite. In
Section 3 we describe the analysis performed to the simulations
outputs in order to search and characterise the overdensities. In
Section 4 we present our results to finally discuss them and present
the conclusions in Section 5.

\section{Numerical Procedures} 
\label{sec:Procedures}
The numerical setup of the $N-$ body simulations used in this work
comprises two stages. In the first instance the galaxies were
generated in isolation, in this case, we generate a host disc galaxy
and a spheroidal satellite galaxy, both with and without gas. We used
these galaxy models to explore different merger configurations. In the
following sections we describe in detail each part of the procedure.

\begin{figure*}[h]
  \begin{center}
    \hspace*{-1cm}\includegraphics[width=0.9\textwidth]{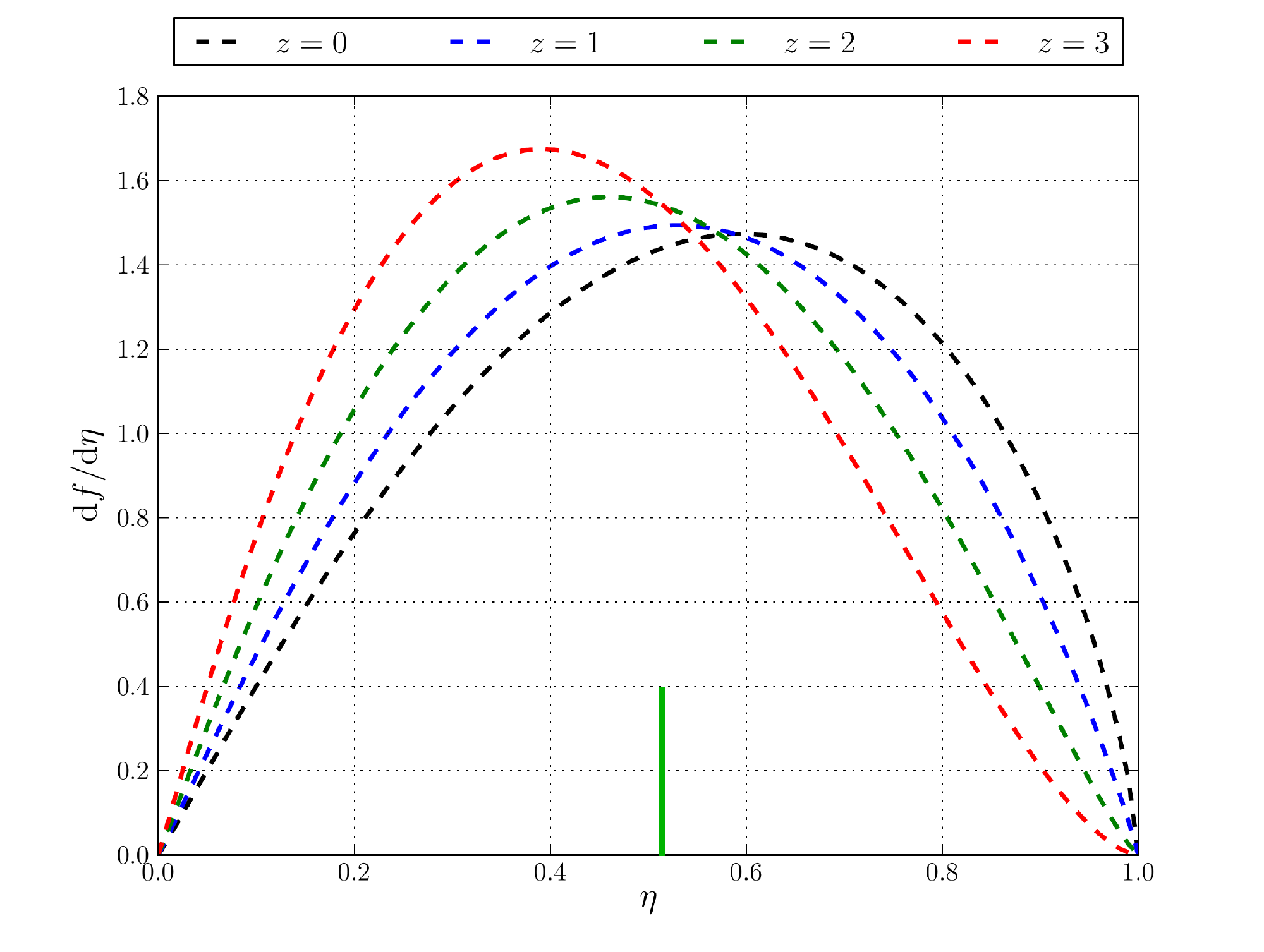}
    \caption{Circularity distribution for the infalling satellites at
      different redshifts. The small vertical line indicates the
      average circularity at $z=2$.}
    \label{fig:circularity}
  \end{center}
\end{figure*}

We used the code Gadget2 to run all our simulations (details of the
code can be found in \citet{Springel2005}). Gadget2 is a general
purpose code to study the evolution of collisionless gravitational
systems. Collisionless particles representing stars and dark matter
evolve only under gravity using a tree method. To follow the evolution
of gas an entropy based smoothed particle hydrodynamics (SPH) scheme
\citep{Springel2002} is used with adaptive smoothing lengths, allowing
conservation of energy and entropy in adiabatic regions. A
synchronization scheme within the integration scheme is used, this is
a quasi-symplectic KDK leap-frog with adaptive individual
time-steps. The code uses a parallelization algorithm based on a
space-filling curve getting high flexibility with high accuracy in
tree force estimation.

\subsection{Initial conditions}
\subsubsection{Isolated Galaxies}
The host galaxy in this work consists of a disk galaxy composed of a
stellar disk and a dark matter halo. Neither gas in the disk nor a
central spheroid is included in the model. The satellite galaxy is
modelled as a spherical galaxy with a collisionless spheroid hosting a
gaseous sphere in hidrostatic equilibrium.

Initial conditions were computed using moments of the collisionless
Boltzmann equation \citep{Hernquist1993, Springel2004}. The dark
matter halo of both galaxies follow a Hernquist density profile with
scale length parameter adjusted to fit the shape of the NFW density
profile as done in~\citet{Springel2005}.

Masses for the galaxies are taken from the CLUES
simulations~\citep{Gottloeber2010, Forero2011}. The mass of the dark
matter halo hosting the disk galaxy is $7.9\times 10^{11}$\hMsun with
a concentration parameter of $c=4.15$. The satellite galaxy has a
total mass of $3.2\times 10^{10}$\hMsun, and $c=4.26$. Since it is not
reasonable to simulate the formation of globular clusters observed
today using properties of current host galaxies, the masses and
properties of these two progenitor galaxies are related to the
properties of the Milky Way galaxy and one of its satellites at $z=2$
as observed from the constrained simulations made by CLUES. Galaxy
disk structure (disk scale length, etc.) is modelled using the
prescription of ~\citet{Mo1998} from which the scale parameters of the
disk are $r_d=1.53$ kpc and $z_0=0.31$ kpc.\\
\begin{figure*}[h]
  \begin{center}
    \hspace*{-1cm}\includegraphics[width=0.9\textwidth]{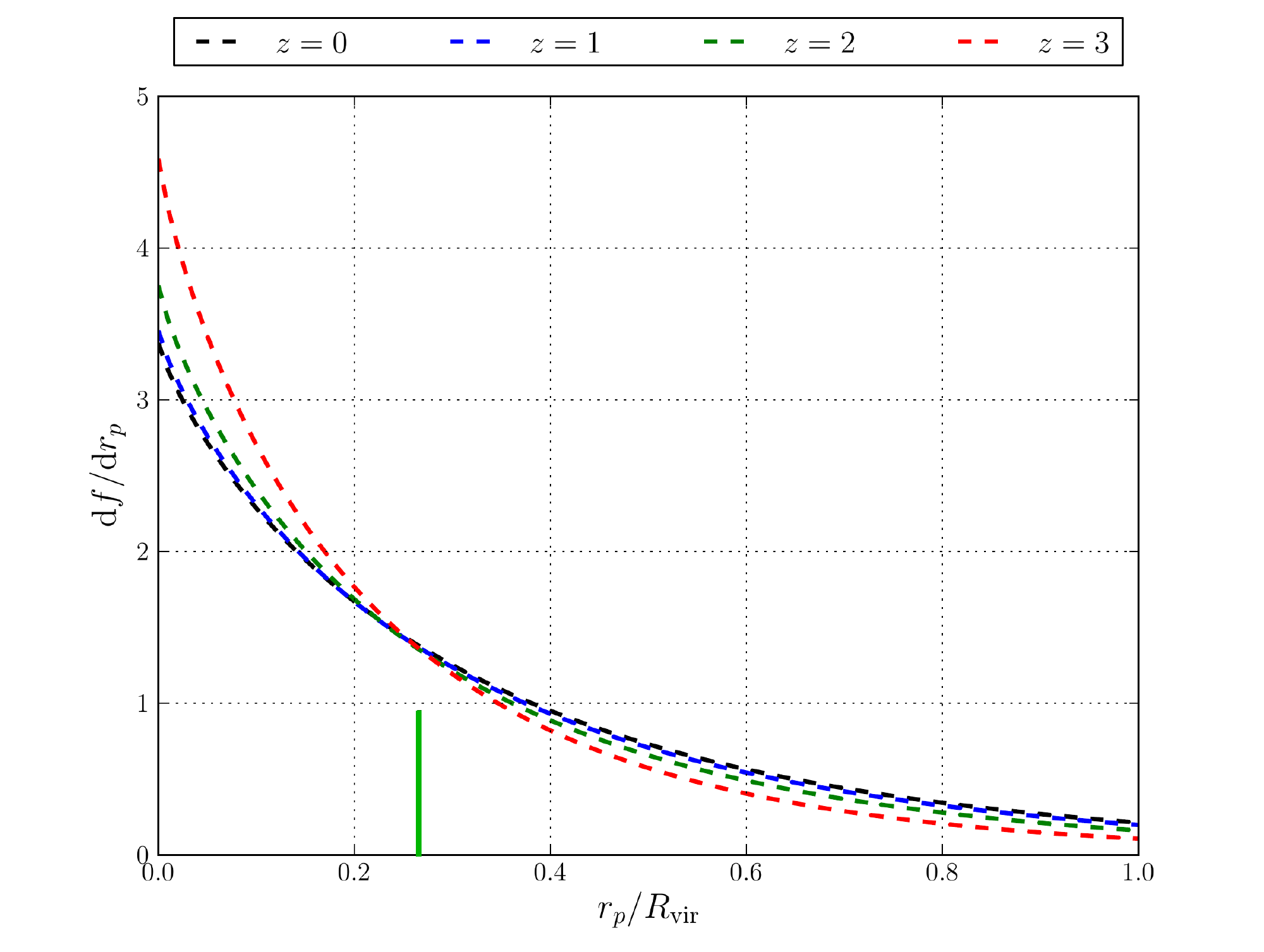}
    \caption{Pericentre distribution for the infalling satellites; the
      redshift dependence is explicitly noted. The small vertical line
      indicates the average pericentric distance value at $z =
      2$. \label{fig:pericentre}}
  \end{center}
\end{figure*}
The host galaxy has a stellar disk with a mass of $3.3\times
10^9$\hMsun where we have used~\citet{Moster2010} to estimate the
total stellar mass for the given dark matter halo at $z=2$ and assumed
that all the stellar mass is deposited in the disk. Since we are not
interested in the evolution of the gas in the disk of the host galaxy
and we assume it does not have a dominant effect on the dynamics of
the merger, we do not include a gaseous component into this galaxy.

The satellite galaxy is composed of collisionless particles
representing dark matter or stars, and has also a gaseous
component. Initially, the gas follows a density profile similar to the
profile of the dark matter halo in hydrostatic
equilibrium. Hydrostatic equilibrium is guaranteed through gas
temperature which is computed as~\citet{Mastropietro2005}

\begin{equation}
  T(r)=\frac{m_p}{k_B\rho_g(r)}\int_r^{\infty}\rho(r)\frac{GM(r)}{r^2}\mathrm{d}r,
\end{equation}

where $m_p$ is the proton mass, $k_B$ is the Boltzmann constant and
$\rho_g(r)$ is the gas mass density. In order to provide a favorable
scenario for the formation of clusters from the material deposited in
the stream, the total gas mass in the satellite has been chosen to be
$\sim 16\%$ of its total mass, providing the scenario for a gas rich
merger. Although arbitrary, the gas fraction is in no way larger than
the cosmological baryon fraction in a dark matter halo
\citep{Lin2008}. We could have included a disk of cold gas in to the
satellite, however this would have implied a new degree of freedom in
our simulations (see next section). Since the direction of the disk
may affect the formation of a stream and formation of potential
candidates to GCs in our simulations, we decided to go for a simpler
spherical distribution looking for a point that is general enough to
study the formation of potential GCs in our simulations. We claim that
if any structure is formed with this setup, for sure, they can be
formed in more favorable conditions where a disk provides cold gas to
the stream. \\
\begin{figure*}[h]
  \begin{center}
    \includegraphics[scale=0.25]{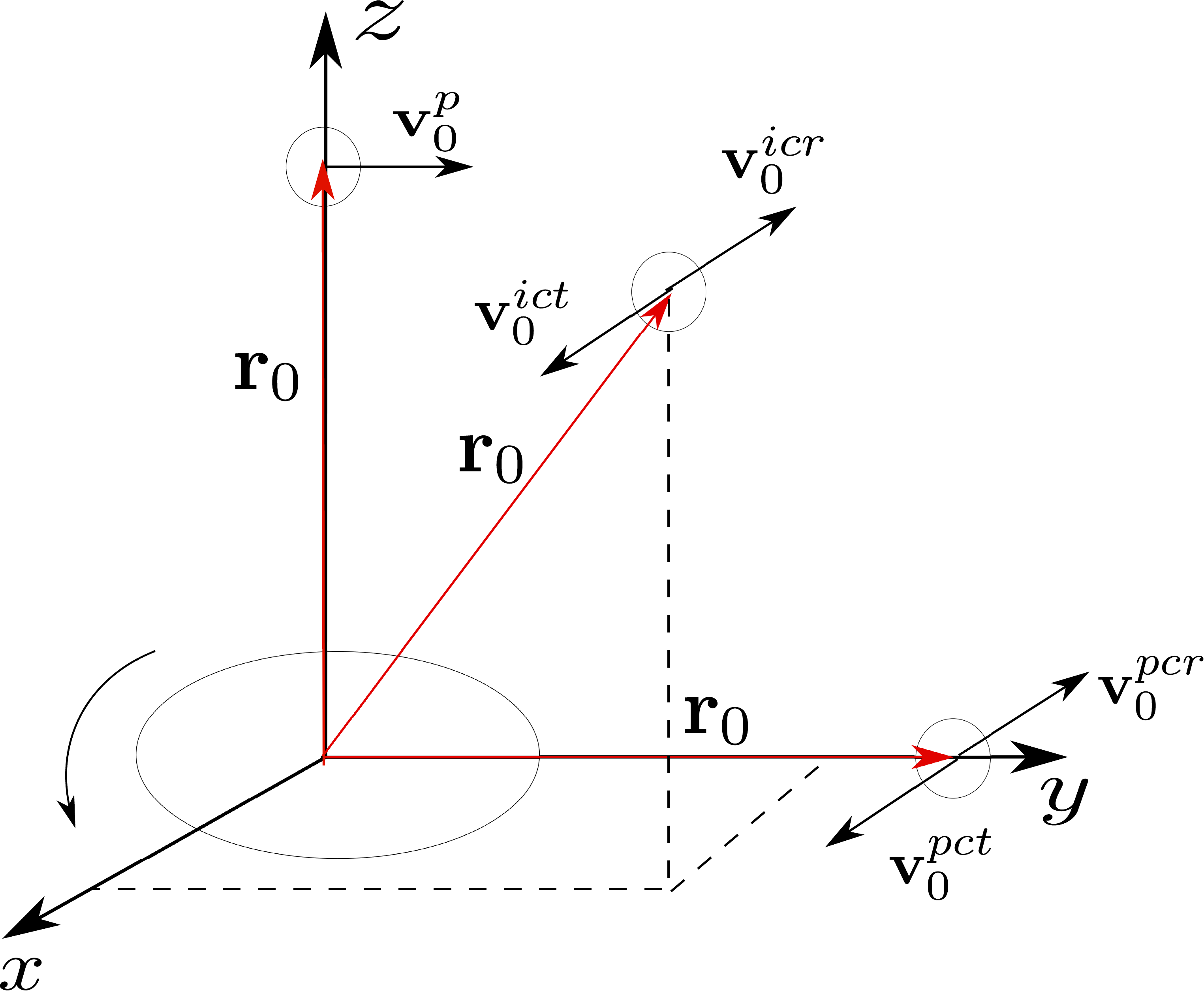}
    \caption{Schematic representation of the initial orbital
      configurations for the original five simulations. The host disc
      is rotating counterclockwise in the $x-y$ plane. $\mathbf{r}_0$
      and $\mathbf{v}_0$ are the initials position and velocity in
      each case. For the inclined configurations, the angle with the
      $z$ axis is $\phi=45^{\circ}$. See table
      \ref{tab:orbitalconfiguration} for details on the coordinate
      values.}
  \end{center}
  \label{fig:orbitalscheme}
\end{figure*}
All galaxies are simulated in isolation after the generation of
initial conditions in order to allow for numerical relaxation of the
initial conditions. Figure \ref{fig:NumRelax} shows, for the dark
matter halos, the convergence of the profiles from the initial
conditions to the final relaxed density profile. Note that the mass
distribution only changes in the very inner region and after the first
1 Gyr the profile is relaxed. Also, the satellite galaxy reaches
relaxation basically very close from the beginning. This check is
relevant since it is important to make sure that there is no numerical
artificial evolution on the density distribution of the galaxies,
since in this way we can ensure that any change in the mass
distribution of the system during the merger is due to the dynamics of
the merger and is not spurious numerical noise or any instability
originated from the initial conditions.

\subsubsection{Merger configuration}

The mergers we plan to study in this work are somehow artificial in
the sense that they do not correspond to the simulation of any
realistic system. However these simulations must reproduce the reality
of our universe. In that sense, there is an infinite set of possible
merger simulations we could run, each with a different orbit. To avoid
running many orbits, and at the same time trying to reproduce the
expected results from our understanding of the universe, we will use
the results shown in \citet{Wetzel2011} to choose the orbits to be
studied in this work. In their work \citet{Wetzel2011} study the
probability distribution of orbital parameters of infalling satellite
galaxies. From them, we use the mean orbital parameters as those of a
representative merger that is in agreement with the current
cosmological paradigm.



Then, to configure the merger we need to obtain realistic values of
the initial position $\mathbf{r}_0$ and velocity $\mathbf{v}_0$ of the
satellite galaxy. For that, from \citet{Wetzel2011}, we use the
circularity $\eta$ and the pericenter $r_p$ distance that depend on
the host halo mass $M_{\mathrm{host}}$ and redshift $z$ and that for
our host halo mass are distributed at the moment of their passage
through the host's virial radius according to the distribution
functions shown in figures \ref{fig:circularity} and
\ref{fig:pericentre}. In both figures, the mean values of the
circularity and the pericentre at $z=2$ are highlighted with a small
vertical green line.

Orbit circularity has a nearly constant small rate of decrease with
redshift while pericenter distance exhibits a decrease in its average
values with $z$. In particular, at $z = 2$ we obtain an average
pericentric distance of $0.27R_{\mathrm{vir}}$, with
$R_{\mathrm{vir}}$ the virial radius of the host halo. For this halo
$R_{\mathrm{vir}}\approx r_{200} = 63.29$ kpc. The average circularity
at $z = 2$ is 0.54. With this two values we calculate the eccentricity
$e$ and apocentric distance $r_a$ using the two body approximation as

\begin{equation}
  e = \sqrt{1-\eta^2},  \\ 
\end{equation}
\begin{equation}
  r_a = \left(\frac{1+e}{1-e}\right)r_p. 
\end{equation}

For our system, the numerical values were found to be $e = 0.84$ and
$r_a = 198.34$ kpc. Finally, making use of the \textit{vis-viva}
equation, the velocity at apogalacticon is simply

\begin{equation}
  v_a = \sqrt{2\frac{GM}{r_a}(1-e)},
\end{equation}

\begin{flushleft}
\begin{figure*}[h] 
\raggedleft
\flushleft
    \includegraphics[width=1.0\textwidth]{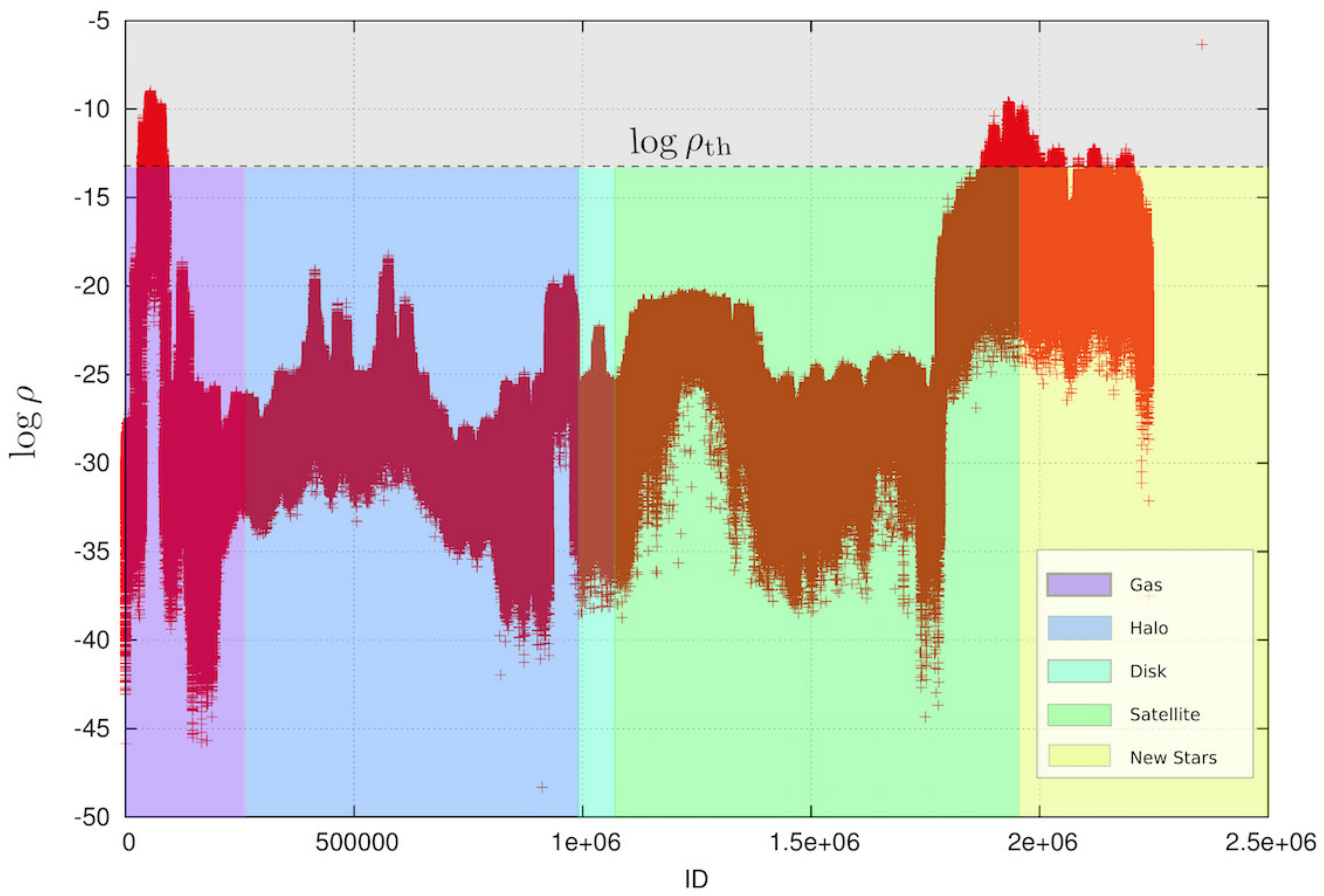}
    \caption{Density values for the particles in GAS2 $p-$simulation
      against the identification particle number. This plot correspond
      to the snapshot at 3.75 Gyr after simulation starts. Density
      units are $10^{10} \mathrm{M}_{\odot}/\mathrm{kpc}^3$}
    \label{fig:gas_density_threshold}
\end{figure*}
\end{flushleft}

which turns to be 34.9 km/s for our infalling satellite.\\

In all simulations the host galaxy disk was always in the $x-y$ plane
with its angular momentum aligned with the z-axis. Based on the
orbital parameters given in the previous paragraph, the merger was
disposed in five different configurations. The only difference between
each configuration is their location relative to the disc plane and
its orbital motion direction relative to the disc rotation. The
configuration parameters are shown in
table~\ref{tab:orbitalconfiguration} and a schematic illustration of
all of them is represented in figure~\ref{fig:orbitalscheme}.

\subsection{Simulations} 
\label{sec:sims}

Some of our simulations include star formation modeled as shown in
~\citet{Springel2003}. In this model a cold gas particle is able to
convert part of its mass in stars when several criteria are met. Its
temperature should be lower than $10^4$ K and its density should be
larger than a predefined threshold density
($\rho_{\mathrm{th}}$). Additionally, the cooling time should be
shorter than the collapse time of the cloud
$t_{cooling}<\frac{1}{\sqrt{G\rho}}$ and the local gas flow must be
negative ($\nabla \cdot \vec{v}<0$). These conditions guarantee that
gas-rich regions, where the star formation process must happen, are
colder, denser and undergoing collapse. Once a gas particle meets
criteria to form stars, they are formed stochastically with a sampling
determined by the local star formation rate. We refer the reader to
~\citet{Springel2003} for further details on the model and
implementation of the star formation.  Our goal is to study the
posibility of formation of globular cluster-like structures in this
kind of mergers. We are not interested in to study the process of star
formation in these candidate structures since it will be very much
dependent on the model and implementation of the different feedback
mechanisms and star formation. Our interest introducing star formation
and feedback in our simulations is to provide the gas with some sort
of realistc conditions that can be meet for candidate structures to
form.

  

\begin{table*}
  \begin{center}
    \begin{tabular}{c c c c}
      \hline
      Name & Nomenclature &  $\mathbf{r}_0$ (kpc) & $\mathbf{v}_0$(km/s) \\[0.1cm]
      \hline
      Perpendicular &  \textit{p}  & (0,0,198.34) & (0,34.9,0)\\
      Planar Corrotating & \textit{pcr} & (0,198.34,0) & (-34.9,0,0)  \\
      Planar Contrarotating & \textit{pct} & (0,198.34,0) &  (34.9,0,0) \\ 
      Inclined Corrotating & \textit{icr} & (99.6,99.6,140.25) & (-24.67,24.67,0)  \\
      Inclined Contrarotating & \textit{ict} & (99.6,99.6,140.25) & (24.67,-24.67,0) \\ 
      \hline
    \end{tabular}
    \caption{Specifications of the satellite's initial orbital
      configurations chosen for each simulation. For a schematic view
      of each configuration, see figure 4. }
    \label{tab:orbitalconfiguration}
  \end{center}
\end{table*}
As it is well known, SPH suffers from fragmentation instabilities that
lead small gas clumps to cluster forming a set of non-physical
structures~\citep{Bate1997, Torrey2013}. Since what we are looking for
in our simulations is exactly fluctuations in the mass distribution we
need to make sure that we find candidates that are not just spurious
numerical fragments formed due to the SPH instability. In order to
avoid this, we ran the same set of initial conditions for several
different particle resolutions and we found that the substructures
present in the lower resolution simulation were recognizable in the
higher resolution simulations, maybe with a little spatial
displacement due to changes in the global dynamics of the system as it
can be seen in figure \ref{fig:candidates-simulations} (See section
\ref{sec:identification} to see how these substructures were
identified). Increasing the resolution of the simulations allow us to
verify that what we find as substructure candidates are true
candidates and not numerical artifacts. Section \ref{sec:Resolution}
describes in better detail the results of our convergence study.

We design two sets of experiments to explore the formation of
substructures in the tidal streams of the satellite galaxy. The first
consists in pure collisionless systems, or in other words, gas-free
simulations. The main purpose of these first experiment was to verify
if the collisionless matter alone could cluster and form bound systems
without the influence of gas. This set of simulations was named DMO
(Dark Matter Only), specifically DMO1 and DMO2 whose only difference
is the number of particles in the satellite as it is shown in
Table~\ref{tab:tab2} where we show the masses, number of particles and
mass per particle of each galactic component in our models.\\

The second set of simulations included gas in the satellite and were
designated with the nomenclature GAS. The difference among them is the
increased resolution, being GAS3 the one with the highest
resolution. Table~\ref{tab:tab2} summarizes the resolution
specifications of the GAS experiment.\\

As it can be seen in table \ref{tab:tab2} the SPH particle mass is of
the order of $5\times10^{3}$ \hMsun \ for GAS3. If we assume that
typical masses for globular cluster candidates are of the order of
$10^{5}$ to $10^{7}$ \hMsun in this simulation we could resolve
globular cluster like structures with between 20 to 2000 gas
particles. Again, we are not interested in to study star formation in
those objects (which will imply the necessity of larger resolution
simulations in order to sample properly star formation inside the
clusters) but study the collapse of gas in the candidate structures,
therefore these numbers are good enough for the purposes of our
work. Finally, we have ran our simulations during a time interval of
the order of 7 Gyr, long enough to study the evolution of the
satellite remnants as it would be observed in present time.

\begin{table*}[h]
\begin{center}
\begin{tabular}{c|c|c|c|c}\hline
\multicolumn{5}{c}{COLLISIONLESS SIMULATIONS}\\
\hline
Name & Component & Mass ($\textrm{M}_{\odot}$)              & $N_p$             & $m_p$ ($\textrm{M}_{\odot}$)  \\\hline
DMO1 & Satellite & $3.2\times10^{10}$  & $1.0\times10^{5}$  & $3.2\times10^{5}$  \\
     & Disk      & $3.3\times10^{9}$   & $5.6\times10^{4}$  & $6.4\times10^{4}$  \\
     & Halo      & $7.9\times10^{11}$  & $7.3\times10^{5}$  & $1.1\times10^{6}$  \\\hline
DMO2 & Satellite & $3.2\times10^{10}$  & $2.0\times10^{5}$  & $1.6\times10^{5}$  \\
     & Disk      & $3.3\times10^{9}$   & $5.6\times10^{4}$  & $6.4\times10^{4}$  \\
     & Halo      & $7.9\times10^{11}$  & $7.3\times10^{5}$  & $1.1\times10^{6}$  \\
\hline
\multicolumn{5}{c}{COLLISIONAL SIMULATIONS}\\
\hline
Name & Component & Mass ($\textrm{M}_{\odot}$)               & $N_p$             & $m_p$ ($\textrm{M}_{\odot}$)  \\\hline
GAS1 & Satellite & $2.5\times10^{10}$  & $4.0\times10^{5}$   & $6.2\times10^{4}$  \\
     & Gas       & $5.0\times10^{9}$   & $2.0\times10^{5}$   & $2.5\times10^{4}$  \\
     & Disk      & $3.3\times10^{9}$   & $5.6\times10^{4}$   & $6.0\times10^{4}$  \\
     & Halo      & $7.9\times10^{11}$  & $7.3\times10^{5}$   & $1.1\times10^{6}$  \\\hline

GAS2 & Satellite & $2.5\times10^{10}$  & $8.0\times10^{5}$   & $3.1\times10^{4}$  \\
     & Gas       & $5.0\times10^{9}$   & $4.0\times10^{5}$   & $1.2\times10^{3}$  \\
     & Disk      & $3.3\times10^{9}$   & $5.6\times10^{4}$   & $6.4\times10^{4}$  \\
     & Halo      & $7.9\times10^{11}$  & $7.3\times10^{5}$   & $1.1\times10^{6}$  \\\hline

GAS3 & Satellite & $2.5\times10^{10}$  & $3.0\times10^{6}$   & $8.3\times10^{3}$  \\
     & Gas       & $5.0\times10^{9}$   & $1.0\times10^{6}$   & $5.0\times10^{3}$  \\
     & Disk      & $3.3\times10^{9}$   & $5.6\times10^{4}$   & $6.4\times10^{4}$  \\
     & Halo      & $7.9\times10^{11}$  & $1.0\times10^{7}$   & $7.9\times10^{4}$  \\
\hline
\end{tabular}
\end{center}
\caption{Collisionless and collisional simulations data. $N_p$ and
  $m_p$ are the number of particles and mass per particle
  respectively.}
\label{tab:tab2}
\end{table*}

\section{Analysis}

\subsection{Density Estimation}
Overdensities are, by definition, regions with a spatial mass density
that is larger than its surroundings. Hence, the best way to identify
them is by estimating the mass density in the body of the tidal
streams. High density regions will be the best candidates to form
autogravitating substructures. We used the EnBiD (Entropy Based Binary
Decomposition) algorithm to calculate the density distribution in real
and phase spaces~\citep{Sharma2006}.\\

The EnBiD algorithm is sensitive to the spatial anisotropies of the
mass distribution by the implementation of the anisotropic smoothing
tensor. In this way, any density underestimation is prevented due to
the ability of the method to use particles along a preferred direction
and not only spherically symmetric around the point of interest, as
the isotropic kernels do. Figure~\ref{fig:gas_density_threshold} shows
the estimated density for the total number of particles in one of the
GAS simulations at $t=3.75$ Gyr. In the figure we show the density of
gas, halo dark matter, disk and satellite particles. New stars formed
from gas particles are also included. Note that this figure only shows
the densities of particles ranked by ID (and type) but allows to see
the high density peaks. As we know, these density peaks are associated
to gravitational instabilities and should be related to anisotropies
in the density distribution of each galactic component, therefore, an
adequate density threshold can be selected to extract the prominent
overdensities in the particle distribution.

As it can be seen in the figure, the overal density of the halo, disk
and a fraction of particles of the satellite have a lower density
value than that for a fraction of particles of gas and stream material
(which in this case corresponds to the last bump composed of satellite
particles and new born stars). The peaks in the values of the
estimated density can be used to fix a density threshold $\rho_{th}$
that can be used to identify global overdensities, as it is shown by
the horizontal line in figure \ref{fig:gas_density_threshold}. Notice
that only a fraction of gas, satellite and new star particles are
above this threshold, and one expects that since those overdensities
are induced by gravitational instability, they are spatially
correlated, as it can be seen in figure
\ref{fig:candidates-simulations}. This density threshold is an
important part of the process of identification of high density peaks
corresponding to the seed of the identification of potential cluster
candidates.

\subsection{Identification of Substructure Candidates}
\label{sec:identification}

Once the densities of the particles have been calculated, we aimed to
identify the overdensities in the field of the stream in order to
label them as possible candidates. We start by estimating density maps
as described in the previous section. These maps highlight the
overdensities above the underlying distribution of particles as it is
shown in figures~\ref{fig:satellite_WO_gas}
and~\ref{fig:satellite_with_gas}. Then, the identification is carried
out by the following series of steps:
\begin{figure*}[h]
  \centering
  \includegraphics[width=0.86\textwidth]{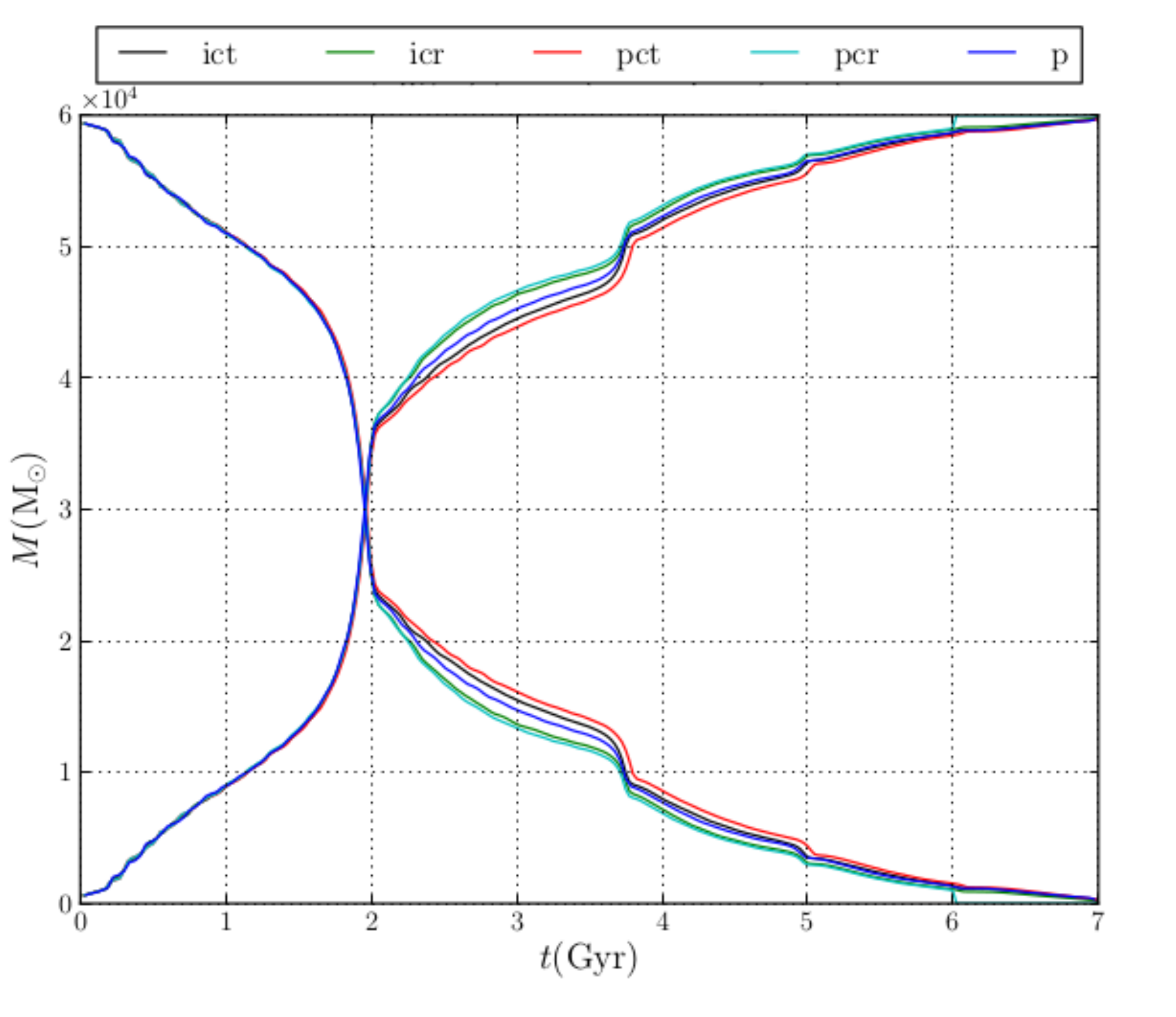}
  \caption{Mass stripped out from the satellite galaxy as a function
    of simulation time represented by the descending curves for each
    orbital configuration. The mass gained by the stream for each
    orbital configuration are the ascending ones.}
  \label{fig:Mstripping}
\end{figure*}

\begin{itemize}
\item First the candidates are identified by performing a selection of
  particles through a phase space density threshold
  $\rho_{\mathrm{th}}$. Particles with phase-space densities below the
  density threshold are ruled out as potential center of some
  candidate clump. The value of $\rho_{\mathrm{th}}$ was chosen
  examining the values of the density of the simulation using, for
  instance, a plot like the one shown in
  figure~\ref{fig:gas_density_threshold} in which we clearly
  distinguish between particles of high and low density. The density
  threshold could be different from one simulation to another.
  
\item Once the particles with $\rho<\rho_{\mathrm{th}}$ are ruled out,
  we elaborate a three dimensional spatial plot of the particles that
  are left. In this plot, it is identified, \textit{by eye} a centre
  of each overdensity. The coordinates of the centre at that
  particular snapshot are then estimated to be $\mathbf{r}_c=(x_c,
  y_c, z_c)$. We chose a random snapshot to do this procedure, but it
  is preferable that the system has had an important evolution, maybe
  after the satellite has passed several times through the disk. The
  fact that at this point we choose by hand the position of the
  candidate has no effect on the results. Using for example a method
  like spherical overdensity would work equally well since we are just
  finding a guess for the position of the overdensity.
  
\item Based on the three dimensional plot built before, it can be
  roughly estimated the size of the overdensity. We assign a spherical
  radius $R_0$, measured from the centre $\mathbf{r}_c$, trying to
  encompass the largest number of overdensity particles. Then,
  particles with position $\mathbf{r}_p=(x_p,y_p,z_p)$, which meet the
  condition $|\textbf{r}_p-\textbf{r}_c|<R_0$ are said to be in the
  first guess of the candidate list. After inspection, we have found
  that using a value of $R_0 \sim 2.0$ Kpc was enough to encompass all
  initial particles in each substructure that will be used later to
  track the actual set of bound particles.
  
 \begin{figure*}[h] 
 \centering
 \includegraphics[width=0.9\textwidth]{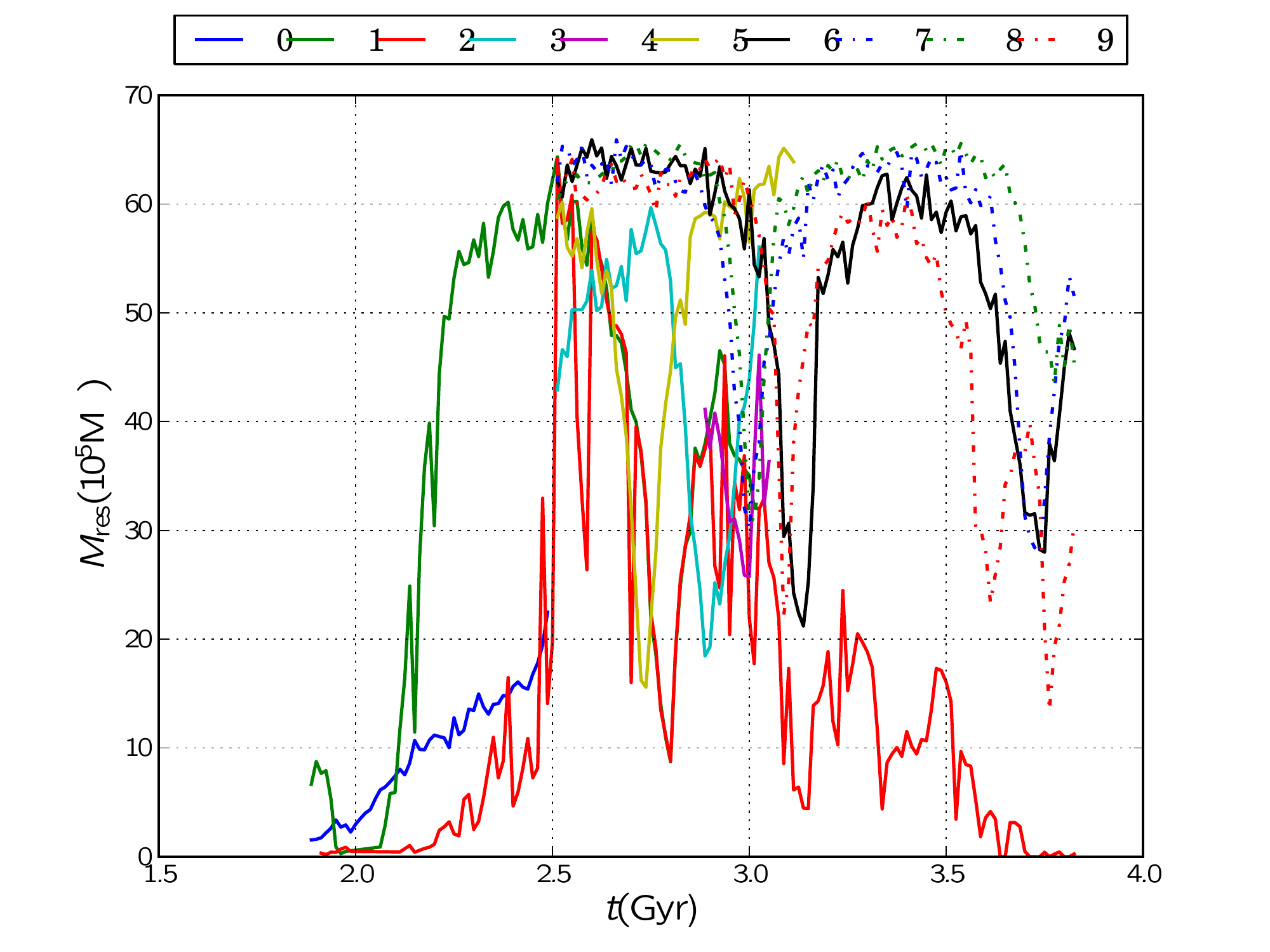}
  \caption{Minimum resolvable masses for the nine candidates in GAS3
    (equation \ref{eq:resolution_mass}). The minimum mass remain much
    smaller than the candidates total mass, see figure
    \ref{fig:candidates-masses}.}
  \label{fig:candidates-minimummass}
\end{figure*}

\begin{figure*}[h]
  \centering
  \hspace*{-1.0cm}\includegraphics[width=1.15\textwidth]{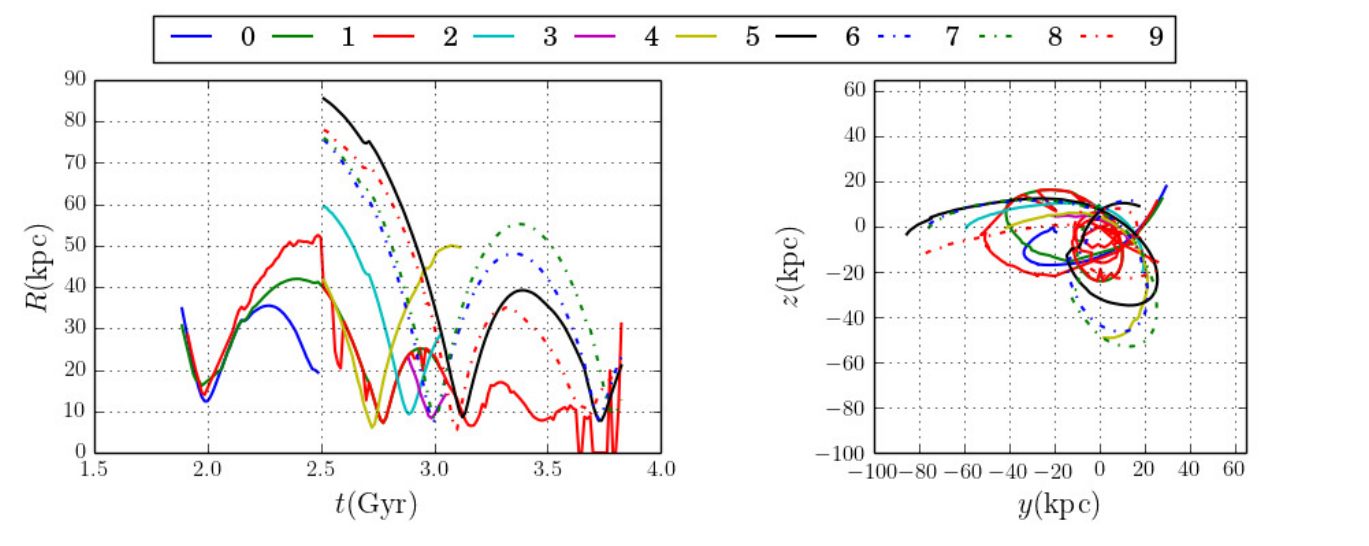}
  \caption{Orbital structure of the nine candidates identified in
    GAS3. \textit{Left panel}, Magnitude of the galactocentric vector
    position as a function of time. \textit{Right panel}, Projection
    of the orbits in the $y-z$ plane.}
  \label{fig:candidates-orbits}
\end{figure*}

\begin{figure*}[h]
  \includegraphics[width=1.0\textwidth]{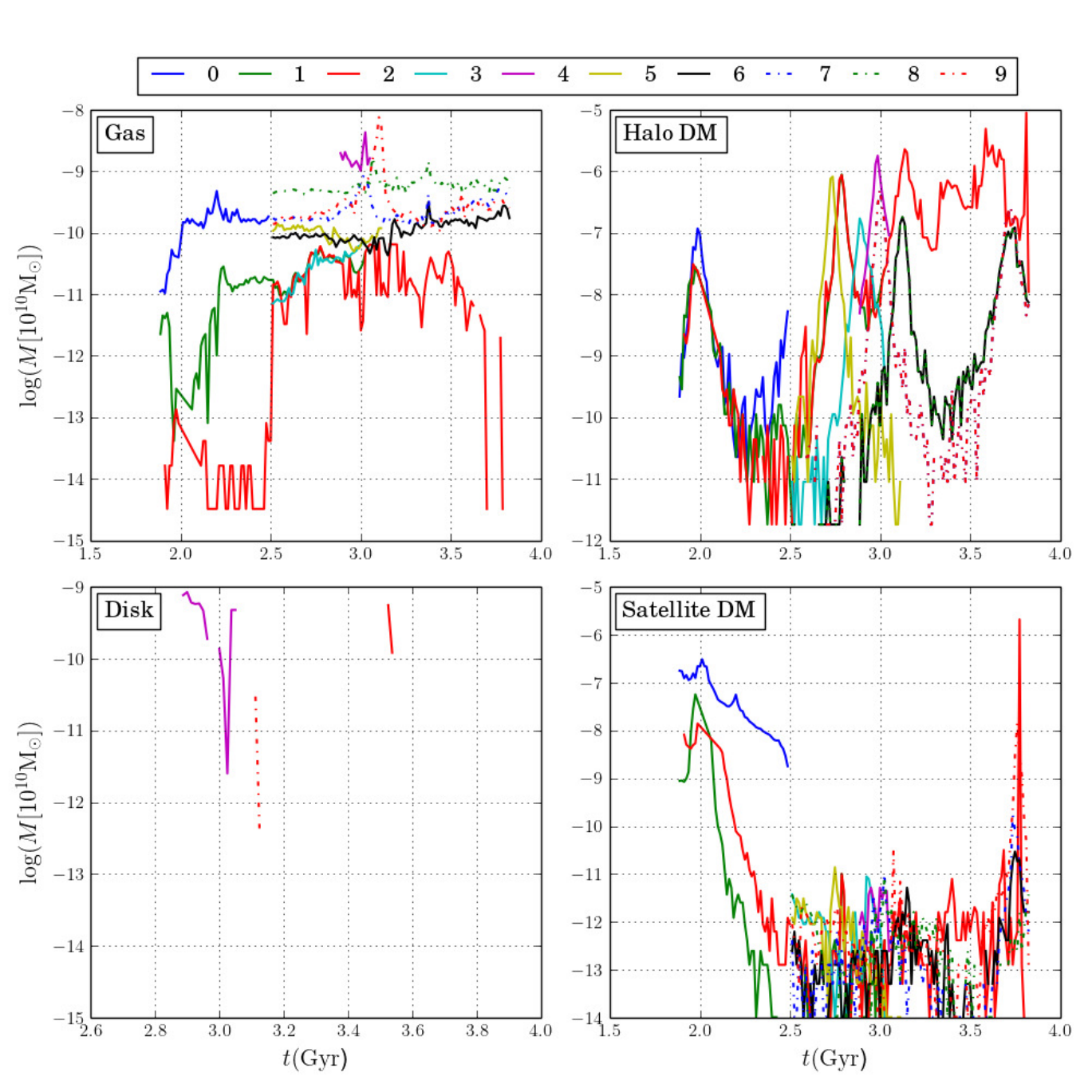}
   \caption{Mass as a function of time for each candidate, segregated
     by type.}
    \label{fig:candidates-masses}
 \end{figure*}

\begin{figure*}[h]
\centering 
\includegraphics[width=0.9\textwidth]{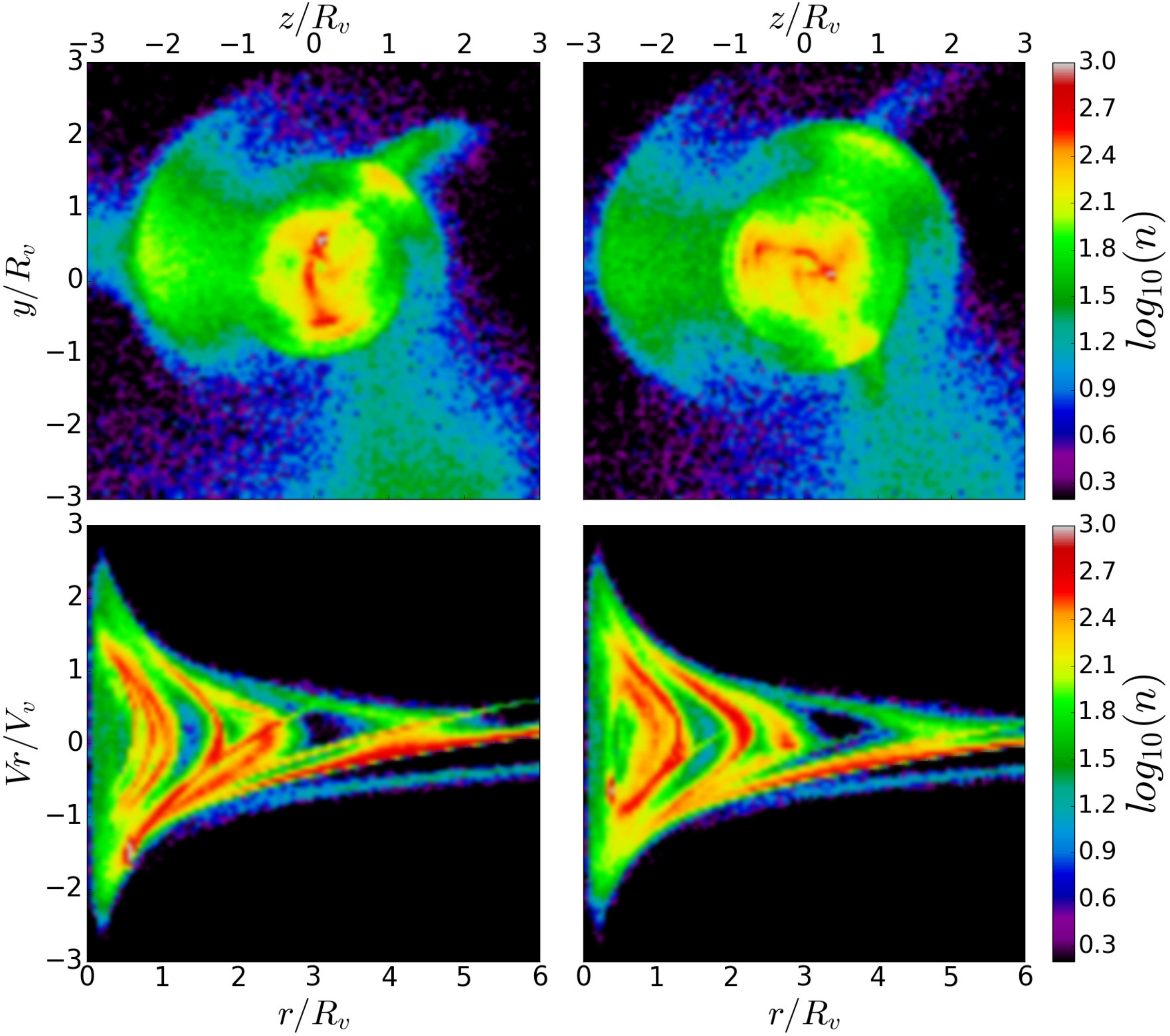} 
\caption{\textit{Top Panels:} Real space projection on the $z-y$
    mapped with density for DMO2 simulation. \textit{Bottom Panels:}
    Phase space projection on the $r-v_r$ plane with same density
    color mapping. The density contrast is high in the the elongated
    radial structures found in the tidal tails, but still do not
    exhibit the morphology of globular clusters. Pannels at the left
    correspond to a time of 6.0 Gyr while right panels are at 7.0
    Gyr. Color values correspond to number density of points in log
    scale.}
  \label{fig:satellite_WO_gas}
\end{figure*}
  
\item With the ID number of each particle in the candidate list, we
  track the position and velocities of such particles in all the
  snapshots in the simulations. At this point, we compute the center
  of mass of the particles in the candidate (for each snapshot) and
  look for particles of any kind that lie within a sphere of radius
  $R_{th}=0.7$Kpc, including dark matter particles from the host and
  the satellite halos, gas, disk particles and new stars born during
  the interaction. Notice that across the snapshots particles can come
  in and out of the sphere of $R_{th}=0.7$Kpc in a way that the list
  of particles that actually belong to the candidate has to be updated
  dynamically.
  
\item Then, for every snapshot, we compute the properties of the clump
  in order to inspect the evolution of the visually identified clouds
  with an astrophysical observed system. Such properties are center of
  mass, energy binding, total mass, the mass by type of particle,
  central and mean densities, tidal and core radii and the tidal
  heating.
  
\end{itemize}


\subsection{Resolution Study Against Artificial Fragmentation}
\label{sec:Resolution}

The numerical scheme used to simulate the hydrodynamics of the gas
could impact the formation of clumps within the molecular clouds in an
artificial way. The resolution of a SPH simulation involving gravity
is therefore a critical quantity in order to obtain realistic results
from physical process rather than artificially induced mechanisms by
numerical fluctuations. \\
As it is widely known, in SPH the properties of gas particles are
obtained by summing the properties of all the particles that lie
within a sphere with a radius known as the smoothing length $h$. The
smoothing lengths are constrained to contain approximately a number of
particles, called number of neighbours $N_{\mathrm{ngb}}$, in the
sphere of radius $h$. Since the gravitational softening is set equal
to $h$, the mass contained in the sphere can not be roughly equal to
the local Jeans mass, otherwise the collapse is inhibited by the
softening of the gravitational forces. \\

\begin{figure*}[h]
  \centering
  \includegraphics[width=0.9\textwidth]{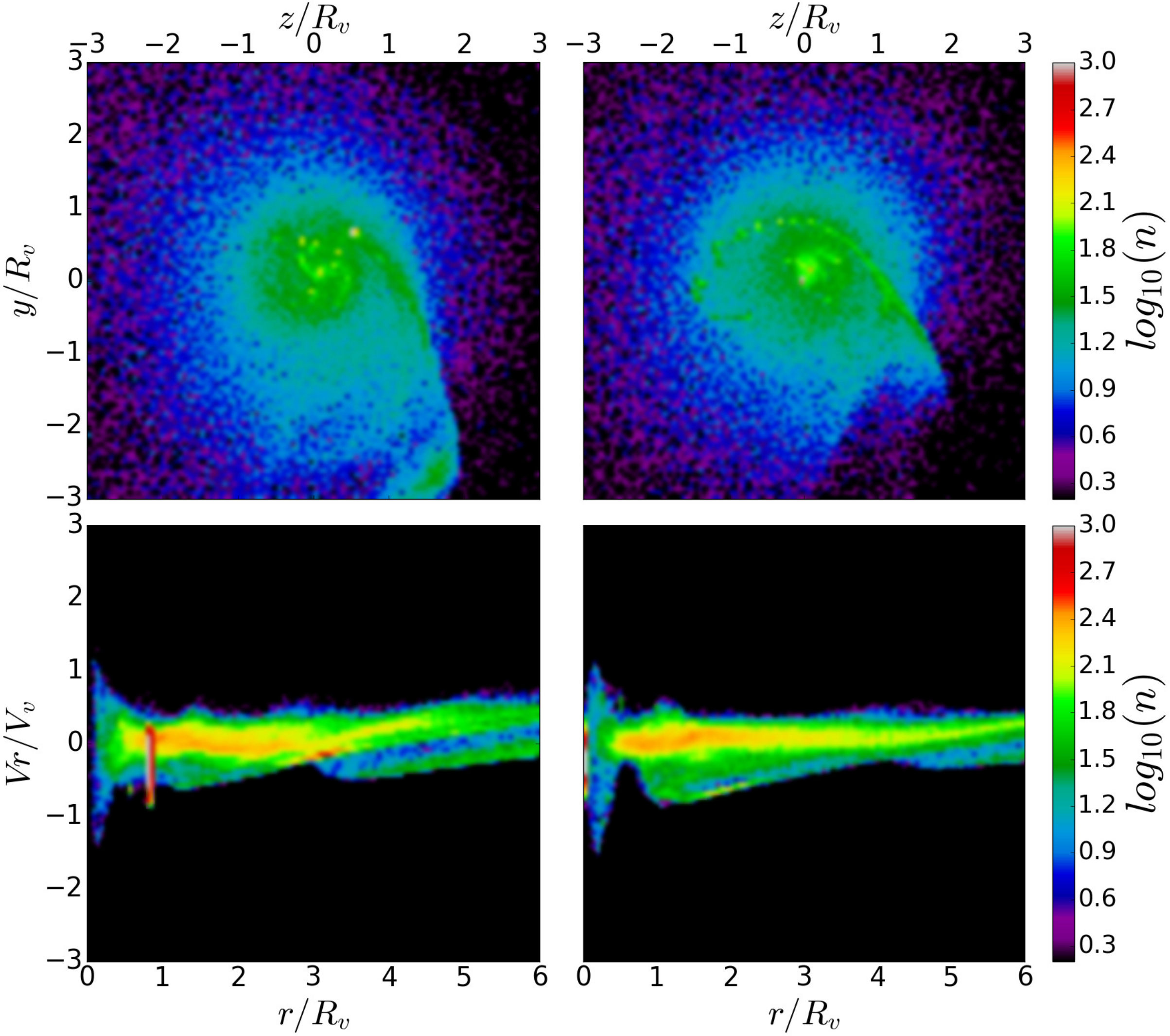}
  \caption{\textit{Top Panels:} Real space projection on the $z-y$
    mapped with density. \textit{Bottom Panels:} Phase space
    projection on the $r-v_r$ plane with same density color
    mapping. $R$ and $V$ are the virial radius and velocity
    respectively. This corresponds to a couple of snapshots of
    GAS2. Panels at the left corresponds to a time of 5.00 Gyr while
    rith panels are at 6.25 Gyr. Color values correspond to number
    density of points in log scale.}
  \label{fig:satellite_with_gas}
\end{figure*}

Thus, the called \textit{minimum resolvable mass}, $M_{\mathrm{res}}$
must always be less than the local Jeans mass $M_J$ given by

\begin{equation}
  M_J=\left(\frac{3}{4\pi \rho} \right)^{1/2}\left(\frac{5k_B T}{\mu
    m_H G} \right)^{3/2},
  \label{eq_jeans_mass}
\end{equation}

where $\rho$ is the density of the gas at temperature $T$, $k_B$ is
the Boltzmann constant, $m_H$ is the mass of the hydrogen atom and
$\mu$ is the gas mean molecular weight~\citep{Draine2011}.  Taking
$M_{\mathrm{res}}$ as the mass of $2N_{\mathrm{ngb}}$ particles, it
can be estimated as~\citet{Bate1997}

\begin{equation}
  M_{\mathrm{res}} =
  M_{\mathrm{gas}}\left(\frac{2N_{\mathrm{ngb}}}{N_{\mathrm{gas}}}
  \right),
  \label{eq:resolution_mass}
\end{equation}

where $M_{\mathrm{gas}}$ and $N_{\mathrm{gas}}$ are the total mass and
particle number of the gas. The previous expression explicitly shows
that for a larger number of particles, the minimum resolvable mass
decreases and the collapse and fragmentation will be less affected for
the numerical implementation. \\

The condition (\ref{eq:resolution_mass}) with $N_{\mathrm{ngb}}=128$
is tested for the clumps in the satellite galaxy gas that we selected
as substructure candidates with the previous recipe. The strategy
adopted for the identification of the progenitors and the results
obtained of such strategy are depicted in the next sections.

Figure~\ref{fig:candidates-minimummass} shows the time evolution of
the minimum resolvable mass for each cluster according
to~\ref{eq:resolution_mass}, which remains much smaller than the local
Jeans mass.



\begin{figure*}[h]
\centering
\begin{center}
\includegraphics[width=0.9\textwidth]{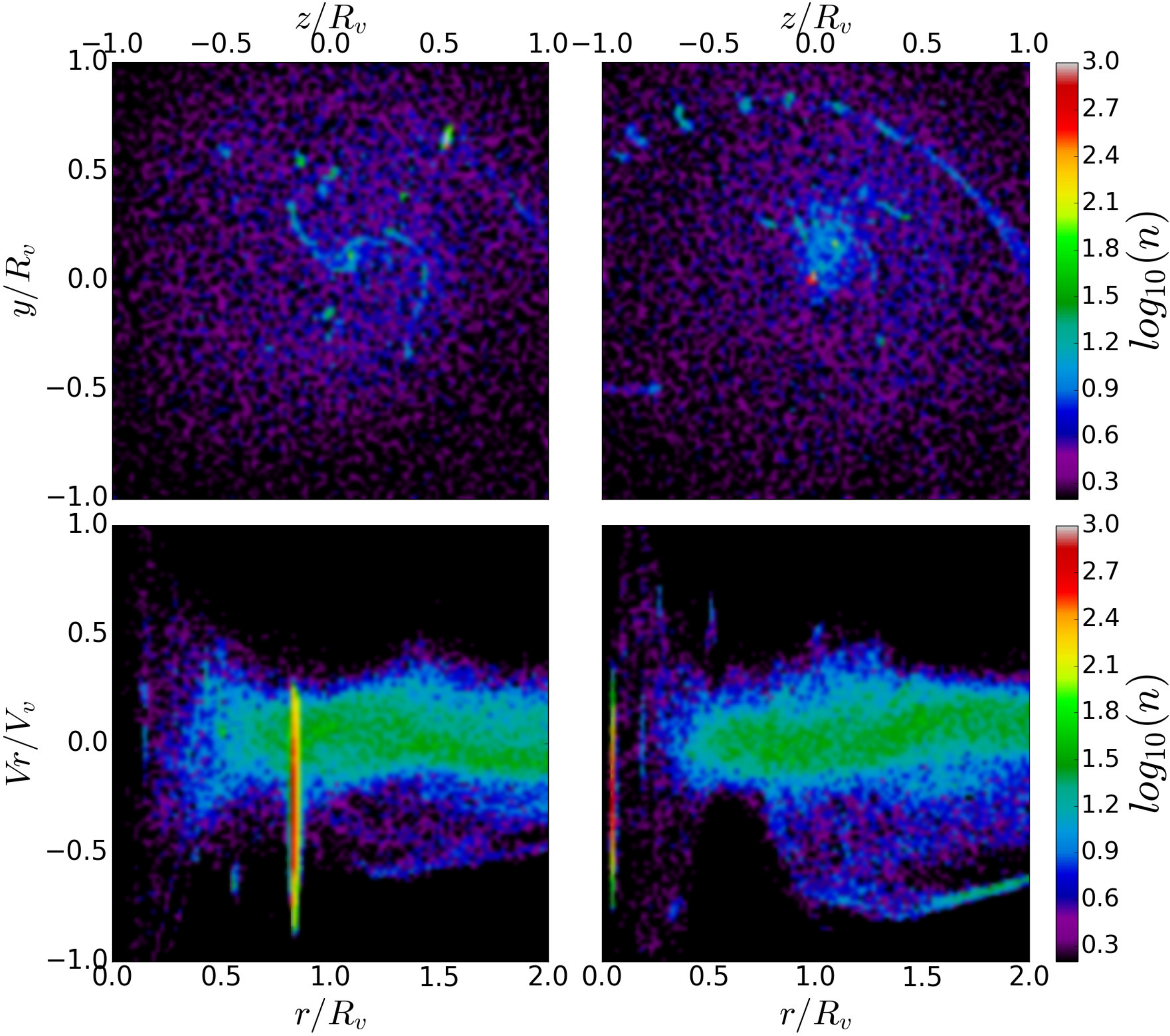}
\caption{\textit{Top Panels:} Real space projection on the $z-y$
  mapped with density. \textit{Bottom Panels:} Phase space projection
  on the $r-v_r$ plane with same density color mapping. This plot is
  exactly figure~\ref{fig:satellite_WO_gas} but zooming to the
  internal region near the galactic disc for GAS2. Panels at the left
  correspond to a time of 5.00 Gyr while panels at the right are at
  6.25 Gyr. Color values correspond to number density of points in log
  scale.}
\label{fig:zoom-gas_density_map}
\end{center}
\end{figure*}

\section{Results}

In figure \ref{fig:Mstripping} we show the mass stripped out from the
satellite galaxy as a function of simulation time. Each line in figure
\ref{fig:Mstripping} represents the evolution of the mass stripped out
from the satellite for each of the five orbital configurations
presented in table~\ref{tab:orbitalconfiguration}.  As it can be seen
in the figure, the rate of mass loss is quite similar for every
orbital configuration of the merger. For this reason, since in our
experiments we found no reason to prefer an orbit from any other, on
the basis of the amount of mass stripped out of the satellite, we
decided, without loss of generality, to run our high resolution
simulations only for the configuration of the orbit perpendicular to
the plane of the galaxy.

\begin{figure*}
  \flushleft
  \hspace*{-1.5cm}\includegraphics[width=1.1\textwidth]{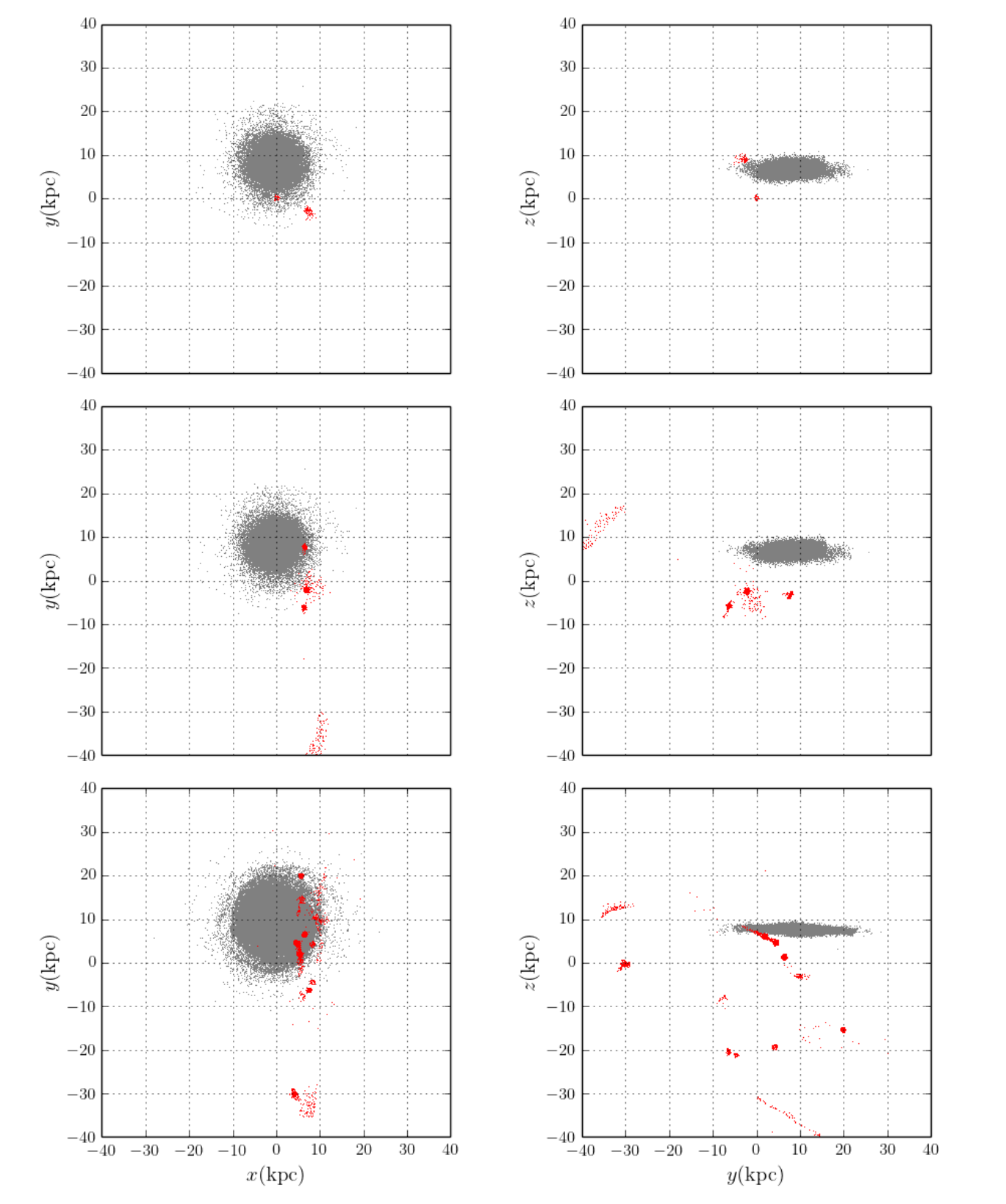}
  \caption{Candidates identified with the algorithm described in
    section \ref{sec:identification}. The number of clumps increase
    with increasing resolution. All plots correspond to 3 Gyr
    simulation time. \textit{Top row}, shows the candidates in GAS1.
    \textit{Middle row}, candidates in GAS2.  \textit{Bottom row},
    candidates in GAS3. In each row, at the \textit{left}, the disc is
    seen face on, and at the \textit{right}, the disc is seen edge on.
  }
  \label{fig:candidates-simulations}
\end{figure*}

As it can be seen in the figure, there are breaks in the mass curve
located at $2,\sim 3.8$ and 5 Gyr. These breaks are associated with
the periastron passages during the merger. Clearly it is the first
passage the one stripping the largest amount of mass out from the
satellite. Most of the mass ejected during the first passage is gas
that is heated up during the collision and should remain bounded to
the host galaxy potentially forming overdensities that we are
interested in our study.

Figure \ref{fig:satellite_WO_gas} shows the projected particle
distribution of the DMO2 simulation at two different time snapshots,
$t=6$ Gyr and $t=7$ Gyr. The figure shows (at the top) in color coded
the density the streams where it can be seen the umbrella effects
associated to the distribution of the merger remnant of a satellite
interacting with a massive host galaxy. At the bottom, each figure
shows the pseudo phase space diagrams, where it can be seen the
disturbances in phase space associated to the structures of the
streams and merger remnants. As it can be seen at the two different
time snaps, there is a diffuse structure that appears at the time 6Gyr
(in real space and phase-space) but that after 1Gyr is already washed
out. This happens to all structures observed in the simulations with
only collisionless matter.

Figure \ref{fig:satellite_with_gas} shows the same projected particle
distribution coded with density colour as the figure
\ref{fig:satellite_WO_gas} but for the GAS2 simulation. Unlike the
case exposed in the previous paragraph, as it is clear from the
projected density and phase-space density, there are more
overdensities and they survived for several orbital periods keeping
their structure for a significant lifetime
fraction. Figure~\ref{fig:zoom-gas_density_map} shows a zoom of the
inner region of figure~\ref{fig:satellite_with_gas} near the galactic
disc. From this result, constrained by the resolution of our DMO
simulations, we conclude that to form long lasting structures we need
cold gas that helps to keep particles bounded gravitationally.

GAS simulations are thus the ones with better results in the formation
of stream substructures. Consequently, we targeted them to apply the
algorithm for identification of substructure candidates whose results
are shown in figure~\ref{fig:candidates-simulations} where the
galactic disc of the host galaxy is also shown as a reference. The
plots show the candidates identified in the simulations GAS1, GAS2 and
GAS3 for the same simulation time of 3 Gyr for comparison purposes. It
is clearly evident that the higher resolution simulation has the
greater number of substructures, which in turn, have the highest
number of particles among all the simulations of this work. For this
reason we only study the properties of the substructures of GAS3.

For each substructure identified in the simulation we investigated
several properties. In GAS3 were identified 10 overdensities
associated to the 10 most densest peaks that we label with numbers
from 0 to 9 and for each one of them we start by determining their
orbital evolution. Figure~\ref{fig:candidates-orbits} (b) shows the
$y-z$ projection of the orbit of the center of mass followed by each
candidate. Figure~\ref{fig:candidates-orbits} (a) shows the distance
between each candidate to the center of the disk of the host galaxy as
a function of time. The more notable aspect of this plot relies in the
fact that the candidates persist among a significant amount of time,
with life times that are of the order of 1Gyr or longer.

Figure~\ref{fig:candidates-masses} shows the evolution of the mass
content of each candidate in GAS3. For all the candidates, the
principal constituent is gas. The high peaks of host dark matter
content present in the candidates are circumstantial particles that
are counted by the algorithm when the candidate traverses the central
region of the dark halo where the density is sufficiently high to
cause the miscounting of host dark particles as candidate
particles. The masses found in each candidate correspond to the masses
measured for globular clusters and high-velocity clouds, both ranging
from 10${}^{3} \mathrm{M}_{\odot}$ to a maximum of 10${}^{6}
\mathrm{M}_{\odot}$ in average, although there are several cases of
clusters with masses above of the 10${}^{6}\mathrm{M}_{\odot}$ value
\citep{Harris1999,Wakker1997}.

\subsection{Dark Matter in Cluster Candidates}

The candidate labeled as Candidate 0 was the only formed by gas and
particles of another species. Figure~\ref{fig:candidates-masses} (a)
clearly shows that the predominant mass component is the dark matter
of the satellite from where it comes. This dark matter component is
not circumstantial, and is an important part of this candidate during
its lifetime. The rest of the candidates are basically cores of gas,
without dark matter or disc stars. This suggest that through this
formation mechanism one could expect to find dark matter in globular
cluster.

\section{Summary and discussion}

In this work we used $N$-body simulations of satellite galaxies
undergoing minor merger with a larger host galaxy. Our goal is to find
if there is formation of globular cluster-like systems in the tidal
stream formed by the tidally stripped material from the satellite. The
work was divided in two main parts: The first part was performed to
explore the possibility of formation of structures from pure
collisionless simulations, the second part was dedicated to simulate
the formation of cluster-like structures from mergers that included
gas.

Then we performed several estimations in the simulations to identify
the stream and the possible autogravitating substructures inside
it. The approach adopted to identify substructures was the estimation
of the phase-space density which reveal the presence of substructures
as density peaks.

The density estimation clearly identifies overdensity regions in which
a cluster-like structure could be formed. As a first conclusion we
argue that without gas, the substructures that could be formed (if at
all) have a sort life as none of the overdensities show a definite
morphology or stability over time. When the gas was included, several
clumps appear.

Running with gas physics results are remarkably different. The
candidates identified in the simulation proved real physical
structures that lived for a considerable amount of time and whose
orbital evolution leads them to be objects in the surroundings of the
galactic disk. The total absence of stars formed within the clumps is
mainly due to the thermodynamic setup of the gas as an initially
isothermal sphere, the temperature of the gas is high enough that
inhibits instantaneous star formation in the candidates. Another
factor at play is the implementation/parameters of the feedback we
used in the simulations that made the effect of feedback to be a bit
to strong in the satellite galaxy. \citep{Oppenheimer2006} discusse
that the original implementation of \citep{springel2003} does not work
equally well for all halo masses and the model and the parameters
should be somehow mass dependent. As it was already mentioned, due to
all the physics involved in the problem of star formation in this kind
of complex scenarios, it is out of the scope of this work to study
star formation in these candidate structures.



The main conclusion of this work is that substructures (globular
clusters and high velocity clouds) could be formed in tidal streams of
gas rich satellites. The validity and scope of this main conclusion
should be tested by running simulations with higher resolutions and
taking in to account different feedback and star formation
models. This is the road map for future work that contributes to
improving and supplementing the results presented here.

Certainly our experiments are limited. First of all, we could study
all other orbital configurations to complement the study. However,
having found that one of the merger configurations already produced
the formation of candidate clusters, the goals of our work where
already meet. Studying under which conditions (merger orbits) it is
more easy to form this kind of structures is an interesting idea, but
indeed it would require a larger amount of computing time in order to
run a suite of simulations to develop the idea.

We could have included gas in to the disk of the host galaxy, we could
also have included a hot gas halo in to the main galaxy (we did not do
it because it made the simulations much more expensive). Both aspects
are important because these gaseous components could affect the
dynamics of the merger and the dynamics of the gas in the stream
through ram pressure stripping and shock heating. Both processes could
work stripping more gaseous material from the satellite, making larger
the amount of gas in the stream, so, we expect it will contribute to
increase the fraction of gas that could end up falling in to cluster
candidates. However the limited design of our experiment shows to
suffice to answer the question on the formation of candidate globular
cluster structures in this kind of processes and we expect that
including those other gaseous components are not going to change the
main conclusions of the work.

\section*{Acknowledgments}
Research work was supported by COLCIENCIAS (doctorados nacionales,
convocatoria 617 de 2013) and the project 111571250082 (convocatoria
715-2015). N.I.L acknowledges financial support of the Project
IDEXLYON at the University of Lyon under the Investments for the
Future Program (ANR- 16-IDEX-0005). Adiotionally, simualtions
performed in this work were run in the computer facilities of GFIF in
the Instituto de Física, Universidad de Antioquia, Hipercubo in the
Instituto de Pesquisa e Desenvolvimento (IP\&D-Univap),
Leibniz-Institut Für Astrophysik Potsdam. The authors gratefully
acknowledge the Gauss Centre for Supercomputing
e.V. (www.gauss-centre.eu) for funding this project by providing
computing time through the John von Neumann Institute for Computing
(NIC) on the GCS Supercomputer JURECA at Jülich Supercomputing Centre
(JSC). Finally, L.F.Q. \& D.A.N. thank Mario Sucerquia for his
meaningful comments in the preparation of the paper.




\begin{thebibliography}
\bibitem[Ashman \& Zepf (1992)]{Ashman1992} Ashman, K., \& Zepf, S.\ 1992, \apj, 384, 50
\bibitem[Bate \& Burkert (1997)]{Bate1997} Bate, M.~R., \& Burkert, A.\ 1997, \mnras, 288, 1060
\bibitem[Bekki \& Chiba (2002)]{Bekki2002} Bekki, K., \& Chiba, M.\ 2002, \apj, 556, 245
\bibitem[Bekki \& Freeman (2003)]{Bekki2003} Bekki, K., \& Freeman, K.\ 2003, \mnras, 346, L11
\bibitem[Belokurov et al. (2006)]{Belokurov2006} Belokurov V., et al., 2006, \apj, 642, L137
\bibitem[Binney \& Tremaine (2008)]{Binney2008} Binney, J. \& Tremaine, S. Galactic Dynamics,\ 2008, Princeton University Press
\bibitem[Binney et al. (2009)]{Binney2009} Binney, J., Nipoti, C. \& Fraternali F.\ 2009, \mnras, 397, 1804 
\bibitem[Blitz et al. (1999)]{Blitz1999} Blitz L., et al.\ 1999, \apj, 514, 818
\bibitem[Carroll \& Ostlie (2006)]{Carroll2006} Carroll, B.~W., \& Ostlie, D.~A, An Introduction to Modern Astrophysics, \ 2008, Addison-Wesley 
\bibitem[Conroy et al. (2011)]{Conroy2011} Conroy C., et al.\ 2011, \apj, 741
\bibitem[Draine (2011)]{Draine2011} Draine, B.~T. Physics of the Interstellar and Intergalactic Medium, \ 2011, Princeton University Press  
\bibitem[Elmegreen \& Efremov (1997)]{Elmegreen1997} Elmegreen, B.~G., \& Efremov, Y.~N.\ 1997, \apj, 480, 235
\bibitem[Forbes \& Bridges (2010)]{Forbes2010} Forbes, D. \& Bridges, T. \ 2010, \mnras, 404, 1203
\bibitem[Forero-Romero et al. (2011)]{Forero2011} Forero-Romero, J.~E., et al.\ 2010, \mnras, 417, 1434
\bibitem[Fraternali et al. (2015)]{Fraternali2015} Fraternali F., et al., 2015, \mnras, 447, L70
\bibitem[\protect\citeauthoryear{Georgiev et al.}{2010}]{Georgiev2010} Georgiev I.~Y., Puzia T.~H., Goudfrooij P., Hilker M., 2010, MNRAS, 406, 1967
\bibitem[Gottloeber et al. (2010)]{Gottloeber2010} Gottloeber, S., Hoffman, Y., \& Yepes, G.\ 2010, arXiv:1005.2687 
\bibitem[Harris (1998)]{Harris1998} Harris, W.~E. Globular Clusters Systems, \ 1998, Springer
\bibitem[Harris (1999)]{Harris1999} Harris, W.~E.\ 1999, 10th Canary Islands Winter School of Astrophysics: Globular Clusters, 325 
\bibitem[Hernquist (1993)]{Hernquist1993} Hernquist L. 1993, \mnras, 86, 389
\bibitem[Ibata et al. (2001)]{Ibata2001} Ibata R., et al., 2001, \apj, 551, 294
\bibitem[K\"upper et al. (2012)]{Kupper2012} K\'upper, A.~H.~W., Lane, R.~R.,\& Heggie, D.~C.\ 2012, \mnras, 420, 2700 
\bibitem[Li et al. (2004)]{Li2004} Li, Y., Law, M., \& Klessen, R.\ 2004, \apj, 614, L29
\bibitem[\protect\citeauthoryear{Lin, et al.}{2008}]{Lin2008} Lin L., et al., 2008, ApJ, 681, 232
\bibitem[Mastropietro et al. (2005)]{Mastropietro2005} Mastropietro, et al. \ 2005, \mnras, 363, 509 
\bibitem[Mo et al. (1998)]{Mo1998} Mo, H.~J., Mao, S., White, S.~D.~M.\ 1998, \mnras, 295, 319
\bibitem[Monaghan (1992)]{Monaghan1992} Monaghan J. 1992, Annual Review of Astronomy and Antrophysics, 30, 543
\bibitem[Moster et al. (2010)]{Moster2010} Moster et al.\ 2001, \apj, 710, 903
\bibitem[Norris \& Kannapan (2011)]{Norris2011} Norris, M.A. \& Kannapan S.J. \ 2011, \mnras, 414, 739
\bibitem[\protect\citeauthoryear{Oppenheimer \& Dav{\'e}}{2006}]{Oppenheimer2006} Oppenheimer B.~D., Dav{\'e} R., 2006, MNRAS, 373, 1265
\bibitem[Peebles \& Dicke (1968)]{Peebles1968} Peebles, P.J.E, \& Dicke R.H. \ 1968, \apj, 154, 891
\bibitem[Price-Whelan et al. (2018)]{Price-Whelan2018} Price-Whelan A., et al \ 2018, Submitted to \apj
\bibitem[Reina-Campos et al. (2019)]{Reina-Campos2019} Reina-Campos M., et al \ 2019, \mnras, 486, 5838-5852
\bibitem[\protect\citeauthoryear{Shapiro, et al.}{2010}]{Shapiro2010} Shapiro K.~L., Genzel R., F{\"o}rster Schreiber N.~M., 2010, MNRAS, 403, L36 
\bibitem[Sharma \& Steinmetz (2006)]{Sharma2006} Sharma S. \& Steinmetz, M.   \ 2006, \mnras, 373, 1293
\bibitem[Springel (2005)]{Springel2005} Springel V. \ 2005, \mnras, 364, 1105
\bibitem[Springel et al. (2004)]{Springel2004} Springel V., White, S.~D.~M. \& Hernquist L. \ 2004, International Astronomical Union Symposium, 421
\bibitem[Springel \& Hernquist(2003)]{springel2003} Springel, V., \& Hernquist, L.\ 2003, \mnras, 339, 289
\bibitem[Springel \& Hernquist (2002)]{Springel2002} Springel V. \& Hernquist, L. \ 2002, \mnras, 333, 649
\bibitem[Torrey et al. (2013)]{Torrey2013} Torrey, P., et al.\ 2013, ASP Conference Proceedings, 477, 237
\bibitem[Wakker \& van Woerden (1997)]{Wakker1997} Wakker, B.~P., \& van Woerden, ~H.\ 1997, Annual Review of Astronomy and Astrophysics, 35, 217
\bibitem[Wetzel (2011)]{Wetzel2011} Wetzel, A.~R. \ 2011, \mnras, 412, 49
\bibitem[Zepf \& Ashman (1993)]{Zepf1993} Zepf, S. \& Ashman, K.\ 1993, \mnras, 264, 611




  
\end{thebibliography}
\end{document}